\documentstyle[prl,aps,multicol,psfig]{revtex}

\begin{document}
\draft

\title{Equilibrium Properties of Temporally Asymmetric Hebbian Plasticity}

\author{Jonathan Rubin$^{*}$,
Daniel D. Lee$^{\dag}$
and H. Sompolinsky$^{\ddag}$}
\address{
$^{*}$Department of Mathematics, Ohio State University, Columbus, OH 43210\\
$^{\dag\ddag}$Bell Laboratories, Lucent Technologies, Murray Hill, NJ 07974\\
$^{\ddag}$Racah Institute of Physics and Center for Neural Computation,
Hebrew University, Jerusalem 91904, Israel}

\date{July 22, 2000, revised: September 8, 2000}

\maketitle
\begin{abstract}
A theory of temporally asymmetric Hebb (TAH) rules which depress or
potentiate synapses depending upon whether the postsynaptic cell fires
before or after the presynaptic one is presented.  Using the
Fokker-Planck formalism, we show that the equilibrium synaptic
distribution induced by such rules is highly sensitive to the manner
in which bounds on the allowed range of synaptic values are imposed.
In a biologically plausible multiplicative model, we find that the
synapses in asynchronous networks reach a distribution that is
invariant to the firing rates of either the pre- or post-synaptic
cells.  When these cells are temporally correlated, the synaptic
strength varies smoothly with the degree and phase of synchrony
between the cells.
\end{abstract}

\begin{multicols}{2}

Recent experimental evidence indicates that synaptic modification in
cortical neurons depends on the precise temporal relation between pre-
and postsynaptic firing \cite{Markram,Bi}.  Presynaptic spikes that
precede postsynaptic firing lead to synaptic potentiation, while those
that follow postsynaptic firing elicit synaptic depression.  The
temporal window for inducing these changes is on the order of 10~msec.
Several recent theoretical studies addressed the potential
implications of this temporally asymmetric Hebbian (TAH) synaptic
plasticity on learning \cite{Abbott,Rao,Levy,Kempter,Eurich,Kistler}.  The
present study is motivated by recent work by Abbott and coworkers who
applied TAH learning in a large population of excitatory presynaptic
cells asynchronously driving a single postsynaptic cell \cite{Abbott}.
Their simulations showed that the distribution of synapses converged
to a bimodal distribution.  The synapses were either almost zero or
had values close to their upper limit.  Moreover, when the firing rate
of the presynaptic cells was increased, the number of strong synapses
decreased so that there was very little change in the output rate.
Thus, the TAH rule seems to provide a mechanism for keeping the mean
output rate invariant.  Since Hebb rules are presumed to underlie
many developmental and learning processes in neuronal systems, it is
important to understand the equilibrium properties of networks with
TAH plasticity and how they depend upon the particular implementation
of these rules.

In this Letter, we study the TAH rule using Fokker-Planck theory
\cite{Levy,Kistler}.  Surprisingly, we find that the behavior of the
system depends crucially on how the boundaries on the allowed range of
synaptic efficacies are incorporated.  In particular, the salient
features found in \cite{Abbott} are unique to an additive learning
rule in which the magnitude of the update does not explicitly depend
on the current value of the synapse.  A very different behavior is
found with a multiplicative rule where the magnitude of the update
decreases as either the upper or lower bounds are approached.

TAH plasticity is described as a change to the synaptic efficacy $w$
between two cells.  A pair of spikes in the input cell and the output
cell, at times $t_i$ and $t_o$, respectively, induces a change in $w$:
\begin{equation}
\label{rule1}
\Delta^\pm w=\pm \lambda f_\pm(w) K(|t_o-t_i|).
\end{equation}
The weight $w$ is increased by $\Delta^+ w$ when $t_o>t_i$ and
decreased by $\Delta^- w$ when $t_i>t_o$.  The temporal dependence of
the update is defined by the filter $K$ which for simplicity is taken
to be $K(t) \equiv \exp(-t/\tau)$.  The coefficient $\lambda$
sets the scale of the synaptic change at each update.  The factors
$f_{\pm}(w)$ determine the relative magnitude of the changes in the
positive and negative directions.

We consider two particular examples of these update rules.  The first
is an additive update rule where the magnitude of the changes is
independent of $w$, so that:
\begin{equation}
\label{add}
f_+=1;\; f_-=\alpha.
\end{equation}
The parameter $\alpha > 0$ denotes a possible asymmetry between increasing
and decreasing the synaptic efficacy.  If the update results in a
synaptic weight outside the bounds $0 < w < 1$, the weight is clipped
to the boundary values.  For the second example, which we will call
the multiplicative rule:
\begin{equation}
\label{multiple}
f_+(w)= 1-w;\; f_-(w)=\alpha w.
\end{equation}
This results in a synaptic increase (decrease) whose magnitude scales
linearly with the distance to the upper (lower) boundary, similar 
to the model in \cite{Kistler}.

To evaluate the equilibrium properties of these rules, the firing
activity in the two cells needs to be specified.  We consider
the case when the input and output activity are stationary stochastic
processes.  The firing of the input cell is characterized by an
instantaneous rate function $\nu_i(t)=\sum_{t_i} \delta (t-t_i)$,
where $t_i$ are the spike times of the input cell with mean rate
$\langle \nu_i \rangle = r_{in}$.  Similarly, the activity of the output
cell is given by $\nu_{o}(t)=\sum_{t_o} \delta (t-t_o)$ with mean rate
$r_{out}$.  The correlation between these two spike trains is described by
the normalized time delayed crosscorrelation function
$C(t)=\left\langle \nu_i(t^\prime)\nu_o(t^\prime +t) \right\rangle/r_{in}
r_{out} - 1$.  Note that for uncorrelated spike trains, $C(t) = 0$.

In the limit of small step sizes ($\lambda << 1$), Eq.~(\ref{rule1})
can be averaged in order to describe the behavior of $w$ on times of order
$1/\lambda$ as a continuous random walk, similar to the approach in \cite{Levy} (see also
\cite{Kistler}).  This random walk has a
mean drift:
\begin{equation}
\label{drift}
v = \langle{dw \over dt}\rangle = r_{in} r_{out} \left[ (f_+-f_-)\tau + f_+T_+ - 
f_-T_- \right]
\end{equation}
where the weighted correlation times, $T_{\pm}$, are 
\begin{equation}
\label{taupm}
T_{\pm}= \int_0^{\infty} dt K(t)C(\pm t).
\end{equation}
The first term in Eq.~(\ref{drift}) is the contribution from
uncorrelated firing activity in the cells and is proportional to $\tau =
\int_0^{\infty} dt K(t)$.  The other terms represent the contribution from the 
synchrony between the two spike trains and are 
proportional to $T_{\pm}$.  
The expression for the diffusion constant $D(w)$ of this random walk is more
complex and will be presented elsewhere. Here we note that 
$D$ is small since it is proportional to the $\lambda$. 
According to Fokker-Planck theory, the equilibrium density
$P(w)$ can be described as a Gibbs distribution with a plasticity
potential $U(w)$ where in the limit of small $\lambda$:
\begin{equation}\label{pot}
U(w) \equiv -\lambda \log[P(w)]
\approx - 2 \int_0^w dw^\prime v(w^\prime)/D(w^\prime) 
\end{equation}
Thus, $P(w)$ will be concentrated near the global minima of $U(w)$.
Depending upon the implementation of the model, the minimum can be
located at an interior point where the drift $v(w)$ vanishes, or at
the boundaries $w=0$ and $w=1$.  To evaluate whether the distribution
of $w$ contains a peak at $0<w<1$ or at the boundaries, the
specific form of the correlation function $C(\pm t)$ needs to be
considered.

We first consider the simple example where the spike train $\{t_i\}$
is a homogeneous Poisson process and the output spike train is a
shifted version of the input train, i.e., $t_o=t_i+\Delta t$ where
$\Delta t$ is the temporal shift between the two spike trains.  In
this case, $r=r_{in}=r_{out}$, and $C(t)=r^{-1}\delta(t-\Delta t)$, and
$T_{\pm}=r^{-1}\exp(-|\Delta t|/\tau)\theta(\pm\Delta t)$, where $\theta(x)=1$
if $x>0$ and $0$ otherwise.
For the additive model in Eq.~(\ref{add}), this leads to a net drift:
\begin{equation}
\label{adddrift}
v = \left\{ \begin{array}{cc}
 (1-\alpha)\tau r^2 + r e^{-\Delta t/\tau}, & \Delta t > 0 \\
 (1-\alpha)\tau r^2 - \alpha r e^{\Delta t/\tau}, & \Delta t < 0.
\end{array}
\right.
\end{equation}
Here $v$ (as well as $D$) is independent of $w$ and the potential
$U(w)$ is $U(w) \approx -2vw/D$. In the limit of small $\lambda$,
the equilibrium distribution will be a $\delta$-function centered at
$0$ when $v<0$ and at $1$ if $v>0$.

\begin{figure}
\leavevmode\centering\psfig{file=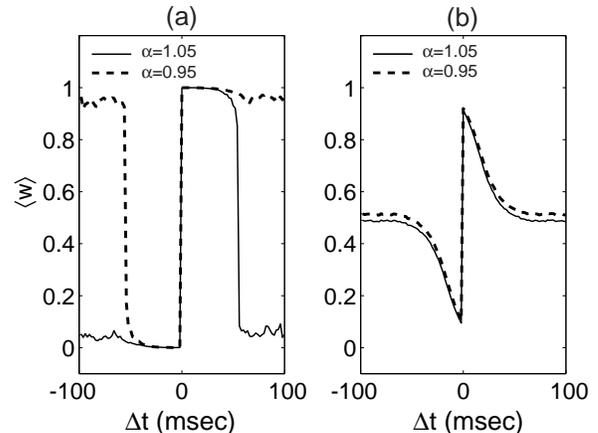,width=3.25in}
\caption{Simulation results showing 
the mean synaptic efficacy $\langle w \rangle$ of a single
synapse undergoing (a) additive and (b) multiplicative TAH
plasticity. $\Delta t$ is the time delay imposed on the spike times of
the post synaptic cell relative to those of the pre synaptic ones.
All simulations are with $\tau=10$~msec and
$\lambda=0.005$.}
\label{fig:single}
\end{figure}

These results are confirmed by the simulations shown in Figure 1(a) and 2(a), 
where we have taken $\alpha =
1.05$ and $0.95$.  For $\alpha = 1.05$, the magnitude of the negative
change is slightly larger than the positive one.  The mean synapse is
zero except when $0 < \Delta t < \Delta t_0$ with $\Delta t_0 \approx
50$~msec.  In this range, the positive correlation between the input
and output cells overcome the negative bias in the update rule to
generate a positive drift so that $w \approx 1$.  The transition at
$\Delta t_0$ is precisely the point where $T_+ = (\alpha - 1)\tau$, see Eq.
(\ref{adddrift}).
This behavior is highly sensitive to whether
$\alpha$ is larger or smaller than 1.  For $\alpha = 0.95$, the mean
synapse is at zero only in the range $-\Delta t_0 < \Delta t < 0$
where the negative correlation is larger than the positive bias.
Otherwise $w\approx 1$.

In contrast, for the multiplicative model the drift velocity is given by: 
\begin{equation}\label{multdrift}
v = \left\{
\begin{array}{ll}
\left[1-(1+\alpha)w\right] \tau r^2 + r (1-w) e^{-\Delta t/\tau}, & \Delta t > 0 \\
\left[1-(1+\alpha)w\right] \tau r^2 - \alpha w r e^{\Delta t/\tau}, & \Delta t < 0
\end{array}
\right.
\end{equation}
Here the drift depends on $w$. It is positive for small $w$ and 
becomes negative for large values of $w$.
In this case, $U$ has an approximately parabolic shape with a minimum
located at $w=w_0$ where the drift velocity vanishes:
\begin{equation}
\label{wmult}
w_0 = \left\{
\begin{array}{ll}
 1-\alpha \left[1+\alpha + (\tau r)^{-1} e^{-\Delta t /\tau}\right]^{-1}, & \Delta t>0 \\
 \left[1 + \alpha(1+(\tau r)^{-1} e^{\Delta t/\tau})\right]^{-1}, & \Delta t<0.
\end{array}
\right.
\end{equation}
This leads to a distribution $P(w)$ with a narrow peak at $w_0$, as
shown in Fig.~2(b).  For large values of $\Delta t$, the input and
output cells are essentially uncorrelated, in which case
$w_0=1/(1+\alpha) \approx 0.5$ for $\alpha \approx 1$.  For positive
$\Delta t \lesssim 50$~msec, the positive correlation between the two
cells gives rise to a mean $\langle w \rangle > 0.5$ as shown in
Fig.~1(b).  Conversely, for small negative $\Delta t$, the reverse
corelation leads to a mean $\langle w \rangle < 0.5$.  Thus, through
this learning rule, the synapse smoothly encodes the temporal phase
relationship between the presynaptic and postsynaptic cells.  A
similar dependence is found when one varies the degree of synchrony 
between the two cells rather than its phase.

\begin{figure}
\leavevmode\centering\psfig{file=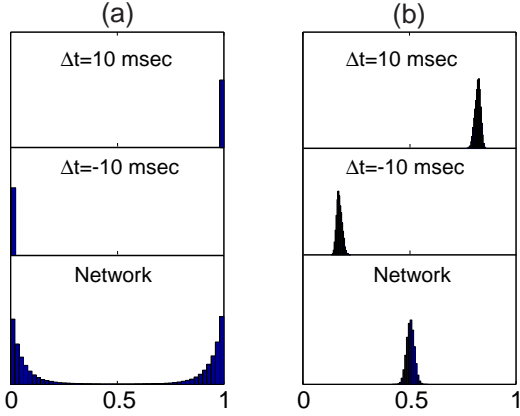,width=2.75in}
\caption{ Histogram showing the distribution of synaptic efficacies
using the (a) additive and (b) multiplicative TAH rule with
$\tau=10$~msec, $\alpha=1.05$, $\lambda=0.005$.  The upper two
histograms in each graph shows the behavior of a single synapse as
described in Fig.~1. The lowest histograms are the distribution of
synaptic efficacies in a network of $N=1000$ excitatory cells with
Poisson activation times and mean input rate $r_{in} = 10$~Hz.  These
cells converge on a single integrate and fire cell with parameters
$\tau_m=20$~msec, $\tau_s=5$~msec, $V_s=5$, and $g_s=0.01$.}
\label{fig:hist}
\end{figure}

Let us now consider the situation where a large population of $N$
cells with modifiable excitatory synapses $w_i$ drive a single
postsynaptic cell.  In the numerical simulations below, the output
neuron obeys dynamics commonly known as ``Integrate and Fire'', where
the potential of the cell is described by the equation: $\tau_m
\dot{V} = - V - I_{s}$.  $\tau_m$ is the passive time constant of the
cell, and when the potential $V$ reaches the threshold $V=1$ it is
reset to zero.  $I_{s}(t)$ is the synaptic current generated by the
$N$ excitatory cells.  Each spike in the presynaptic cells triggers a
contribution to the output cell's conductance that decays with
synaptic time constant $\tau_s$, yielding:
\begin{equation}
I_{s} (t) = g_{s} \sum\limits_{i=1}^N w_i (t)  
\sum\limits_{t_i<t}
e^{(t_i-t)/\tau_s)} (V-V_{s}) .
\end{equation}
$V_s$ is the reversal potential for the excitatory synapses.  The
synaptic efficacy $w_i(t)$ describes the increase in the output cell's
conductance in units of $g_s$, immediately after a spike in the $i$-th
cell.  The peak conductances, $w_i$, are in turn modified by the TAH
dynamics described above.

\begin{figure}
\leavevmode\centering\psfig{file=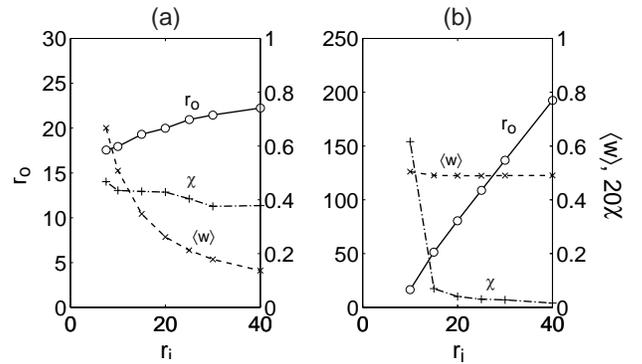,width=3.25in}
\caption{The effect of increasing the input rate, $r_{in}$ on the
equilibrium state of a network of $N=1000$ excitatory synapses driving
an integrate and fire cell, undergoing (a) additive and (b)
multiplicative TAH dynamics. Shown are the output firing rate
$r_{out}$, the mean
synaptic efficacy $\langle w \rangle$, 
and the correlation strength $\chi$.  The latter was determined by
fixing one of the synapses to 0.5 and numerically integrating
Eq.~(\ref{taupm}) and using Eq.~(\ref{chi}).
}
\label{fig:network}
\end{figure}

Here we present a general theoretical analysis of the system which is
independent of the details of the output cell dynamics.  
In the limit of small $\lambda$ Eqs. (\ref{drift})-(\ref{pot}) holds for 
each of the $i$-th synapses, with its correlation times $T^i_{\pm}$ 
and $T^{'i}_{\pm}$ defined using 
the correlation $C_i(t)$ between the $i$-th cell and the output. 
As before, we
will assume that the inputs are described by independent homogeneous
Poisson processes, all with mean rate $r_{in}$.  However, the statistics
of firing in the output cell as described by its mean rate $r_{out}$ and
$\{C_i(t)\}_{i=1}^N$, is
determined by its response to the incoming spikes rather than
determined externally as in the previous example.  We first describe
the behavior of the additive model.  
Due to causality and the
Poisson nature of the input spikes, $T^i_{-}=0$.  For $T^i_{+}$, we
make the following plausible assumptions (i) since the output cell
is driven by a large number of asynchronous inputs, $T^i_{+}$ is
positive but small; and (ii) its value increases roughly linearly with the
synaptic efficacy of the presynaptic $i$-th cell, namely,
\begin{equation}
\label{chi}
T^i_{+} \approx \tau \chi w_i, \; \chi>0
\end{equation}
where (iii) $\chi = \chi(r_{out})$ is expected to be a monotonically
decreasing function of the firing rate of the output cell.  $\chi$ is
larger at smaller $r_{out}$ when the output cell spends more time near
threshold and thus is more sensitive to the timing of the incoming
spikes.  Eq.~(\ref{chi}) implies that the drift for each synapse, $v_i
\propto 1-\alpha + \chi w_i$. Thus, if $\alpha - 1$ is positive and of
order $1$, the system will converge to a state where all the $w_i$ are
zero, or when $\alpha < 1$, $w_i \approx 1$.  An interesting situation
occurs when $0<\alpha-1\ll 1$ so that the weak negative bias can
balance the weak positive correlations.  In this case, the diffusion
constant $D$ is approximately constant, and the potential is given by:
\begin{equation}
\label{Uadd}
U(w) \approx {4 \over 1+ \alpha^2 }
  \left[ (\alpha-1 ) w - {1\over 2} \chi w^2 \right]
\end{equation}
which has local minima at the boundaries, $w=0,1$.  This equation has to 
be solved self-consistently since $r_{out}$ which determines $\chi$ is 
itself dependent on $U$. Over a wide range of input rates the 
self-consistent solution is an unsaturated state, in 
which $P(w)$ has significant weight 
both near $0$ and near $1$, which in turn implies that
$U(0)=U(1)$ up to order $\lambda$. Hence by Eq. (\ref{Uadd}), this
state is characterized by an output rate $r_{out}^*$ such that
\begin{equation}
\label{invariant}
\chi (r_{out}^*) =  2(\alpha-1) .
\end{equation}
More precisely, as $r_{in}$ increases, $r_{out}$ increases slightly by an
amount of order $\lambda$ inducing a decrease in $\chi$ of that
order. This leads to a small relative increase in $U(1)$, which in
turn reduces $P(1)$ by an amount which roughly compensates for the
increase in $r_{in}$, maintaining Eq.~(\ref{invariant}). This behavior is
confirmed by simulations whose results are shown in Fig.~2(a) and
3(a).  As $r_{in}$ increases from 10 to 40~Hz the output rate remains
approximately constant at $\approx 22 Hz$, and $\chi \approx 0.1$ in
agreement with eq. (\ref{invariant}). The mean efficacy $\langle w
\rangle$ decreases to compensate for the increase in $r_{in}$, as found in
\cite{Abbott}.

The multiplicative model results in a very different state.  In fact,
since the correlations are weak for large $N$, the equilibrium
behavior is similar to the previous example with only a single input
and output cell with large $\Delta t$.  In particular, like
Eq.~(\ref{wmult}), the potential $U(w)$ has a single minimum at the
point of zero drift:
\begin{equation}
w_0 \approx {1 \over 1 + \alpha }
\end{equation}
Thus, the distribution is highly concentrated near $w_0$, as seen in Fig.~2(b),
and is largely independent of the mean rates of both the input and
output cells.  As $r_{in}$ increases, the output rate also increases and
is similar in behavior to a cell with fixed synapses, $w_i \approx
w_0$.  Note that as $r_{out}$ increases, $\chi$ decreases but this results
in only a small decrease in the mean synaptic efficacy.  Finally, we
note that in contrast to the additive model where the boundaries are
local minima of $U$ and the point of zero drift is a maximum of $U$,
in the multiplicative model $U(w)$ has a single minimum at the point
of zero drift.  Hence, the equilibration time of the additive model
will in general be much slower than that of the multiplicative model
since synapses have to overcome the potential barrier of the synaptic
potential.  Indeed, this difference in equilibration times is seen in
our numerical simulations.

We have shown here that the multiplicative TAH rule leads to a very
different equilibrium distribution of synapses compared with an
additive rule.  Most importantly, the multipicative model is not
sensitive to moderate changes in the parameters of the plasticity rule
and does not suffer from slow convergence.  Furthermore, experimental
results reveal a dependence of the magnitudes of the synaptic changes
on the amplitude of the initial synaptic efficacy, which supports a
multiplicative TAH rule \cite{Bi}. The observed mean {\it fractional}
negative change remained constant over a wide range of synaptic
efficacies, and is consistent with the assumption that the negative
change is proportional to $w$ as described by a multiplicative
$f_-(w)$ in Eq.~(\ref{multiple}).  The fractional positive changes
monotonically decreased with synaptic efficacy and vanish smoothly at
some maximum value, again in qualitaive agreement with the
multiplicative model.  Although the observed shape of the $w$
dependence deviates from the simple linear form of $f_+(w)$ assumed
here, the qualitative properties of the rule are unaffected by this
difference.  In conclusion, networks with the multiplicative TAH rule
in the asynchronous state should display an equilibrium synaptic
distribution that is largely insensitive to the firing rates of the
pre- and post-synaptic cells.  On the other hand, a coherent temporal
modulation of the firing of the inputs to a target cell leads to a
synaptic distribution which encodes the degree and phase of the
synchrony between the cells in a smooth manner. The functional
implications of this behavior in the development and learning
properties of neuronal systems remains to be explored.

JR is supported in part by the NSF grant DMS-9804447.
HS is supported in part by a grant of the Israeli Science Foundation and the 
Israel-USA Binational Science Foundation.  HS and DDL also acknowledge
the support of Bell Laboratories, Lucent Technologies.

\end{multicols}

\end{document}